\documentclass[%
    groupedaddress,
    preprint,
    showkeys,
    showpacs, preprintnumbers,
    amsmath,amssymb,
    aps, pra,
]{revtex4-1}

\usepackage[utf8]{inputenc}
\usepackage[usenames, dvipsnames]{xcolor}
\usepackage{textcomp}
\usepackage{hyperref}
\hypersetup{colorlinks=true, 
            pdfborder={0 0 0},
            linkcolor=MidnightBlue,
            citecolor=MidnightBlue,
            filecolor=MidnightBlue,
            urlcolor=MidnightBlue}
         
\begin{document}

\title{Resolving a discrepancy between experimental and theoretical lifetimes in atomic negative ions}

\author{Tomas Brage}
\email{tomas.brage@fysik.lu.se}
\affiliation{Division of Mathematical Physics, Department of Physics, Lund University, Sweden}

\author{Jon Grumer}
\email{jon.grumer@teorfys.lu.se}
\affiliation{Division of Mathematical Physics, Department of Physics, Lund University, Sweden}

\date{\today}

\begin{abstract}
Recently the lifetime of the excited $^{2}P_{1/2}$-state of S$^-$ was measured to be $503\pm 54$ s (B\"ackstr\"om \textit{et al.} Phys. Rev. Lett. 114, 143003 (2015)). The earlier theoretical lifetime of $436$ s was clearly outside the experimental error bars. To investigate this discrepancy we have performed systematic and large-scale multiconfiguration Dirac-Hartree-Fock calculations for this system. After including a careful treatment of correlation and relativistic effects, we predict a well-converged value of $492$ s for this lifetime, with an uncertainty considerably less than 1\%, thereby removing the apparent conflict between theory and experiment. 
We also show that this result corresponds to the non-relativistic limit in the $LS$-approximation for the M1 transition within this $^2P$ term. We also demonstrate the usefulness of the latter approach for $^2P$ transitions in O$^-$, Se$^-$ and Te$^-$, as well as for analogous M1 transitions within $^2D$ terms in Ni$^-$ and Pt$^-$ ions. 
\end{abstract}




\maketitle

\section{Introduction}
Negative atomic ions remain an interesting and challenging problem to both experimental and theoretical studies \cite{ref_1a, ref_1b, ref_1c, ref_1d}. In the theoretical models we are faced with strong correlation contributions, since these system are far from dominated by a central, nuclear potential, but are rather defined by electron-electron interactions \cite{ref_2}.
On the experimental side, the spectroscopic identification of these systems is extremely challenging since they only have a few bound states - 
often only the ground state. In cases where more than one state is bound, we can 
observe the bound-bound spectrum of these ions. 
However, we are then restricted to single- and few-line spectroscopy which, except in very few cases \cite{ref_1a, ref_1e}, only involve states of the same parity.
To support the experimental determination, accurate theoretical predictions of transition properties are therefore needed.

Halogen-like negative ions have recently been investigated \cite{ref_1a,ref_3,ref_4}. These have ground states of the form $np^5$~$^2P_{3/2}$, and the $np^5$~$^2P_{1/2}$ as the only excited states.
This opens up a bound-bound decay channel $^2P_{1/2}$ $\rightarrow$ $^2P_{3/2}$, which is dominated by a forbidden, magnetic dipole (M1) transition in all 
the cases investigated here. 

A lingering problem \cite{ref_3,ref_4} in these studies seemed to be a disagreement between experimental \cite{ref_1a} and
theoretical lifetimes from multiconfiguration Hartree-Fock calculations \cite{ref_5}, which casts some doubt on the identification of this transition in S$^-$ (where $n=3$). In this work we therefore present a systematic and elaborate theoretical investigation of the $^2P$ transition in S$^-$ based on the relativistic multiconfiguration Dirac-Hartree-Fock (MCDHF) method, and compare with other available experimental and theoretical results. In addition we also present results for this transition in some other negative ions (O$^-$, Se$^-$ and Te$^-$) as well as for the analogous M1 transition in $^2D$ terms of Ni$^-$ and Pt$^-$ ions. 

\section{Theoretical Method}
In the MCDHF method the atomic eigenstate is represented by an atomic state function~(ASF),
\begin{equation}
\label{ASF}
\Psi(\Gamma;J) = \sum_{i=1}^{N}c_i\Phi(\gamma_iJ) \; ,
\end{equation}
where $\Phi(\gamma_iJ)$ are configuration state functions (CSFs), constructed as 
angular-momentum-coupled, antisymmetrized products of one-electron Dirac-orbitals. The sum over CSFs in (\ref{ASF}) is determined by substitutions of one-electron orbitals from one or more reference states, in this case $1s^2 \ldots\, 3p^5$~$^2P_{1/2,3/2}$, to an active set of correlation orbitals. This is increased systematically while monitoring the convergence in order to build effectively complete wavefunctions and obtain accurate results for the physical quantities of interest - which in this case are fine-structure separations and transition rates.  

The construction of the correlation models also relies on the definition of two parts of the electron cloud of the ions - the core and the valence electrons. In the present approach the core is defined as all subshells with $n < 3$, while the valence subshells are defined as those with $n=3$. Depending on the physical quantity under investigation, the most important correlation is often between electrons in the valence subshells, while core-valence and core-core correlation generally is of less importance in atoms and positive ions (note that core-core effects play an important role when calculating e.g. isotope shifts \cite{ref_16}). In spite of the fact that correlation, including core-valence and also core-core, is more important in negative ions, the partition of the electron subshells into two parts is still a useful concept. The correlation models employed in the present work are defined in the following section.

Details on the MCDHF method and its non-relativistic counterpart can be found in the recently published review \cite{ref_17} and book \cite{ref_18} by Fisher et al. as well as in the comprehensive book by Grant \cite{ref_19}. In the present work we use the MCDHF method as implemented in the most recent version of the \textsc{Grasp2k} code by J\"{onsson} et al. \cite{ref_5}.

\section{Results and Discussions}
\begin{table}
\centering
\caption{\label{sulphur_minus} 
Energies ($E$) and rates ($A$) for the $^2P_{1/2}\rightarrow ^2P_{3/2}$, transition in S$^-$, together with lifetime ($\tau$) of the $^2P_{1/2}$ for different correlation models with systematically enlarged active sets of orbitals. The $\delta E$ values show the relative difference between our \textit{ab initio} energies and the experimental value $E_{exp}$ reported by Blondel \textit{et al.} \cite{ref_3}. This value for the energy and values for the lifetime from other calculations and experiments are given in the bottom of the table. The various model acronyms are defined in the text. The \emph{adjusted} values for $A$ and $\tau$ are rescaled to the experimental energy of Blondel \textit{et al.}}
\begin{tabular}{llrrrlc}
\\ \hline\hline
& \multicolumn{3}{c}{ab initio} && \multicolumn{2}{c}{adjusted} \\
\cline{2-4}  \cline{6-7}
& \multicolumn{1}{c}{$E$} &  \multicolumn{1}{c}{$\delta E$} & \multicolumn{1}{c}{$A$} && \multicolumn{1}{c}{$A$} & \multicolumn{1}{c}{$\tau$} \\ 
 Model  & \multicolumn{1}{c}{(meV)}     &  \multicolumn{1}{c}{(\%)} & \multicolumn{1}{c}{(ms$^{-1}$)} && \multicolumn{1}{c}{(ms$^{-1}$)} & \multicolumn{1}{c}{(s)}   \\
\hline
DF            & 63.3024 &  5.59 & 2.3930 && 2.0327 & 491.97 \\
DFBQ          & 60.6233 &  1.12 & 2.1019 && 2.0327 & 491.96 \\
\hline                                               
VV$_3$SD$_3$  & 56.9078 & -5.08 & 1.7386 && 2.0327 & 491.96 \\
VV$_3$SD$_4$  & 59.0948 & -1.43 & 1.9469 && 2.0327 & 491.96 \\
VV$_3$SD$_5$  & 59.3941 & -0.95 & 1.9756 && 2.0327 & 491.95 \\
VV$_3$SD$_6$  & 59.5337 & -0.70 & 1.9905 && 2.0327 & 491.97 \\
VV$_3$SD$_7$  & 59.5715 & -0.63 & 1.9943 && 2.0327 & 491.96 \\
VV$_3$SD$_8$  & 59.5994 & -0.59 & 1.9971 && 2.0327 & 491.96 \\
\hline                                               
CV$_2$SD$_4$  & 61.8510 &  3.17 & 2.2322 && 2.0327 & 491.96 \\
CV$_2$SD$_5$  & 60.6721 &  1.20 & 2.1070 && 2.0327 & 491.96 \\
CV$_2$SD$_6$  & 60.1256 &  0.29 & 2.0505 && 2.0327 & 491.96 \\
CV$_2$SD$_7$  & 60.4645 &  0.86 & 2.0854 && 2.0327 & 491.96 \\
\hline                                               
VV$_3$SDT$_3$ & 56.6467 & -5.81 & 1.6985 && 2.0327 & 491.95 \\
VV$_3$SDT$_4$ & 58.6979 & -2.09 & 1.9079 && 2.0327 & 491.96 \\
VV$_3$SDT$_5$ & 58.9217 & -1.72 & 1.9298 && 2.0327 & 491.96 \\
\hline
$E_{exp}$           & \multicolumn{6}{l}{$59.9507\pm 0.0004$ \cite{ref_3}} \\
$\tau_{exp}$        & \multicolumn{5}{c}{ } & $503 \pm 54$ \cite{ref_1a} \\
$\tau_{analytical}$ & \multicolumn{5}{c}{ } & $491.93^a$ \\ 
$\tau_{others}$     & \multicolumn{5}{c}{ } & $437$ \cite{ref_4} \\
\hline \hline
\multicolumn{7}{l}{$^a$Eq. (\ref{eq:simple_rate}) with $\sigma_{exp}$ from \cite{ref_3}.} \\
\end{tabular}
\end{table}

The result from various correlation models are presented in Tab. \ref{sulphur_minus} and compared with other experimental and theoretical results. The first column defines the theoretical method and correlation model where the DF acronym refers to Dirac-Fock (i.e. when a single CSF defines the ASF in Eq. (\ref{ASF})) while DFBQ to a model which includes also Breit and leading radiative QED corrections (vacuum polarization and self-energy) \cite{ref_6}. The following models represent systematically enlarged correlation models. VV$_3$ denotes models with valence-valence correlation within $n=3$ subshells while CV$_2$ adds additional core-valence correlation contributions with $n=2$ subshells. SD implies single and double substitutions from the reference set (the $np^5\,^2P_J$ CSFs where $n=3$ for S$^-$) and SDT refers to single, double and triple substitutions. 
The subscript number in the end of the model acronym designation denotes the maximum principle quantum number, $n_{max}$, of the orbitals in the active set. The models represent valence-valence and core-valence correlation, and includes single, double and triple substitutions to an active set of orbitals.

We present two results for
the rates - first the \textit{ab initio}, $A_{ab-initio}$, and second an adjusted value, $A_{adjusted}$, where we rescale the rate with experimental energies according to
\begin{equation} \label{eq:2}
A_{adjusted} =\left(\frac{E_{exp}}{E_{ab-initio}}\right)^3 A_{ab-initio}
\end{equation}
In this equation $E_{exp}$ and $E_{ab-initio}$ are experimental and theoretical \textit{ab initio} transition energies, respectively. This approach is based on the non-relativistic limit, where the line strength, $\mathcal{S}$, is energy independent and the rate is simply $A\propto E^3 \times \mathcal{S}$.

It is clear that the adjusted rate, $A_{adjusted}$, and corresponding lifetime is converging fast and does not change as a function of neither the type of correlation model nor its size, $n_{max}$. Even if it might be argued that the transition energy is not fully converged, it is clear that there is little chance that any additional contributions to the transition rate could affect the final value of $A_{adjusted}$.

From a straightforward analytical and non-relativistic model for the M1 transition in the $LS$-approximation, it becomes clear why the convergence of our results is so fast. The M1 is a forbidden transition, but expected \cite{ref_7, ref_8}, since it is not induced by small corrections to the wavefunction. More importantly, the rate of the M1 transition  in this model, does not depend on the radial part of the wavefunctions, but is rather given by~\cite{ref_9}
\begin{eqnarray}
A_{analytical} & = & \frac{64\pi^2e^2a_0^2(\alpha/2)^2\sigma^3}{3h(2J'+1)} \left|\langle LSJ||{\bf J}^{(1)}+\left(g_s-1\right){\bf S}^{(1)}||L'S'J'\rangle\right|^2 \nonumber \\
& \approx & \frac{2.6973 \times 10^{-11} \sigma^3}{2J'+1} \left|\langle LSJ||{\bf J}^{(1)}+\left(g_s-1\right){\bf S}^{(1)}||L'S'J'\rangle\right|^2
\end{eqnarray}
in a general case of fine-structure, M1 transition, where the primed quantities represent the initial, upper state. In our specific case with a $^2P_{1/2}\rightarrow ^2P_{3/2}$ transition, we get
\begin{eqnarray}
A_{analytical} & \approx & \frac{2.6973\times 10^{-11}  \sigma^3 }{2} \left|\langle^2P_{3/2}||{\bf J}^{(1)}+\left(g_s-1\right){\bf S}^{(1)}||^2P_{1/2}\rangle\right|^2
\end{eqnarray}
where $\sigma = E/hc$ is the wavenumber for the transition (in cm$^{-1}$). It can be shown that
\begin{equation}
\left|\langle^2P_{3/2}||{\bf J}^{(1)}+\left(g_s-1\right){\bf S}^{(1)}||^2P_{1/2}\rangle\right|^2 \approx \frac{4}{3}
\end{equation}
in pure $LS$-coupling. Finally, we find that
\begin{equation} \label{eq:simple_rate}
A_{analytical} \approx 1.7982\times 10^{-11}\sigma^3 .
\end{equation}
By using the experimental transition energy $\sigma = 59.9507$ meV$/hc$ = 483.535 cm$^{-1}$ \cite{ref_3} we get $A_{analytical} = 2.0328\times 10^{-3}$ s$^{-1}$, in perfect agreement with our adjusted result and experiment.

\begin{table}[b]
\centering
\caption{\label{other_ions} Experimental wavenumbers for the $^2P_{1/2}\rightarrow ^2P_{3/2}$ transitions ($\sigma_{exp}$), together with lifetimes for the upper $^2P_{1/2}$ level in four different negative ions. Lifetimes are reported from experiment ($\tau_{exp}$), Eq. (\ref{eq:simple_rate}) using $\sigma_{exp}$ values ($\tau_{analytical}$), and from  calculations ($\tau_{calc}$). In addition, we also present values for $^2D_{3/2}\rightarrow \,^2D_{5/2}$ M1 transitions in two selected negative ions.}
\begin{tabular}{llccc}
\\ \hline\hline 
Ion &  \multicolumn{1}{c}{$\sigma_{exp}$ (cm$^{-1})$}& \multicolumn{1}{c}{$\tau_{analytical}$ (s)} & \multicolumn{1}{c}{$\tau_{exp}$ (s)} &  \multicolumn{1}{c}{$\tau_{calc}$ (s)} \\ 
\hline
\multicolumn{5}{c}{$^2P$-terms}\\ 
 O$^-$ & 177.084 \cite{ref_1d} & $1.00 \times 10^4$ && $7.24\times 10^3$ \cite{ref_4}    \\
 S$^-$ & 483.535 \cite{ref_3}  & 492   & $503 \pm 54$ \cite{ref_1a}  & 437 \cite{ref_4}, 492$^a$\\
Se$^-$ & 2278.15 \cite{ref_4}  & 4.70  & $4.78 \pm 0.18$ \cite{ref_1a}   & 4.92 \cite{ref_4} \\
Te$^-$ & 5005.36 \cite{ref_13} & 0.443 & $0.463 \pm 0.008$ \cite{ref_1a} & 0.454 \cite{ref_4} \\
\multicolumn{5}{c}{$^2D$-terms}\\
Ni$^-$ & 1484.8 \cite{ref_14}  &  18.88 &$ 15.1 \pm 0.4$ \cite{ref_14}   &    \\
Pt$^-$ & 9740.9 \cite{ref_15}  & 0.0669&                     & $0.071$ \cite{ref_15}   \\
\hline\hline 
\multicolumn{5}{l}{$^a$our MCDHF results presented in Tab. \ref{sulphur_minus}} \\
\end{tabular}
\end{table}

This simple model can of course also be used for $^2P$ transitions in other negative ions. Andersson \textit{et al.} \cite{ref_4} reported the lifetimes of the single bound excited states in Se$^-$ ($4p^5\;P_{1/2}$) and Te$^-$ ($5p^5\;P_{1/2}$) using a room temperature magnetic ion storage ring. Recently, a breakthrough in ion-beam storage technology at cryogenic temperatures was reported \cite{ref_11,ref_12}. This new technology allows the storage of beams of negative ions for much longer times than what has been possible before, which brings measurements of atomic lifetimes of the order of tens of minutes within reach. This enabled B\"ackstr\"om \textit{et al.} to measure the lifetime of the $3p^5\;^2P_{1/2}$ level to be $503 \pm 54$ s and to measure the lifetimes of the corresponding states in Se$^-$ and Te$^-$ with considerably higher accuracy \cite{ref_1a} than in earlier studies.

One could also apply this analytical model to systems with different terms. 
The $3d^9\,4s^2\,^2D_{3/2}\rightarrow 3d^9\,4s^2\,^2D_{5/2}$ transition in e.g. Ni$^-$ was recently observed by Kami\'{n}ska et al. \cite{ref_14}. An anologous $^2D$ transition for $5d^9\,6s^2$ configurations was observed in Pt$^-$ by Th\o gersen \textit{et al.} \cite{ref_15} (note that no experimental value for the lifetime was reported in this work). In the $^2D$ case the rate is given by
\begin{eqnarray}
A_{analytical} & \approx & \frac{2.6973\cdot10^{-11} \sigma^3  }{4}\left|\langle^2D_{5/2}||{\bf J}^{(1)}+\left(g_s-1\right){\bf S}^{(1)}||^2D_{3/2}\rangle\right|^2
\end{eqnarray}
where the reduced matrix element is 
\begin{equation}
\left|\langle^2D_{5/2}||{\bf J}^{(1)}+\left(g_s-1\right){\bf S}^{(1)}||^2D_{3/2}\rangle\right|^2 \approx \frac{3}{5}
\end{equation}
for pure $LS$-coupled states. From Tab. \ref{other_ions} it is clear that the fact that the rates are well-represented by the analytical expression of Eq. (\ref{eq:simple_rate}), is not unique for S$^-$, but a general case for at least the halogen-like transitions. 

There are two possible sources of the uncertainty in the predicted lifetimes - the experimental uncertainty in the measured transition energy and the deviation from the $LS$-approximation for the negative ions studied here. The former is for S$^-$ estimated to be less than 0.01~{\textperthousand} while the latter, according to the stability of the adjusted value in Tab. \ref{sulphur_minus}, has an uncertainty of less than 0.1~{\textperthousand}. The uncertainty of the predicted lifetime is therefore well within 1 \% for S$^-$, but increasing with nuclear charge for the other systems in Tab.~\ref{other_ions}, due to an increasing deviation from the $LS$-approximation for heavier ions.

\section{Conclusions}
We can conclude that the rate of the fine structure M1 transitions are straightforward to compute for the negative ions considered here, while an accurate determination of the wavelengths is dependent on a careful treatment of correlation. We therefore recommend the use of the measured lifetime of these upper levels as a support for the spectroscopic identification of the corresponding fine-structure transitions. A similar method was proposed and used for highly ionized species in what was labelled "single-line spectroscopy" for silver-like ions, e.g. Tungsten \cite{ref_10}. 

Measurements of the lifetimes of the upper states involved in these transitions could be used to confirm their spectroscopic identification, This could be an important tool since atomic anions often have just a few bound states, and therefore only a few photon signals from bound-bound transitions, thereby making traditional spectroscopic analyses impossible. In addition to this, in cases where the lifetime, but not the transition energy, is known, the latter can be determined from the former by using the simple energy dependence of the lifetime. 

\section{Acknowledgments}
We gratefully acknowledge Dag Hanstorp, Henrik Cederquist and Henning T. Schmidt for helpful comments and fruitful discussions. This work was supported by the Swedish Research Council (VR) under Contract No. 2015-04842. 

\bibliography{sulphurminus_v3.bib} 


\end{document}